# FOURIER ANALYSIS OF BIOLOGICAL EVOLUTION: CONCEPT OF SELECTION MOMENT

ASHOK PALANIAPPAN

Secondary structure elements of many protein families exhibit differential conservation on their opposing faces. Amphipathic helices and β-sheets by definition possess this property, and play crucial functional roles. This type of evolutionary trajectory of a protein family is usually critical to the functions of the protein family, as well as in creating functions within subfamilies. That is, differential conservation maintains properties of a protein structure related to its orientation, and that are important in packing, recognition, and catalysis. Here I define and formulate a new concept, called the selection moment, that detects this evolutionary process in protein sequences. I present a detailed account of its possible applications.

A multiple alignment of the members of the protein family of interest is created. Conservation at each position of the multiple alignment is calculated in one of two ways:

1) by rigorously taking into account the phylogeny of the protein family, and rate variation among sites; 2) by using the Shannon entropy measure:

$$H(i) = -\sum_{j} p_{ij} \ln p_{ij}, \tag{1}$$

where H($i$) is the entropy at position $i$, $p_{ij}$ is the fractional occurrence of residue type $j$ at position $i$, and the sum is taken over all 20 residues. The Shannon entropy is a simplified measure of conservation in this context. In this way we convert the multiple alignment into a one-dimensional function of a numerical conservation score. Now the presence of a periodicity in this one-dimensional function implies a moment in the pattern of substitution in this protein family. In particular, if a peak of periodicity of this function coincides with the period of an α-helix, then there exists a strong periodicity of selection at 3.6 residues. Selection moment is then defined as the modulus of the Fourier transform of the conservation function, and the selection moment for a given periodicity is given by:

$$S(\theta) = \left| \sum_{n=1}^{N} H(n) e^{i\theta n} \right|, \tag{2}$$

where ($2\pi/\theta$) is the periodicity, n is the residue index, N is the length of the sequence, and radian measure is used. A dominant signal at $\theta = 100°$ reflects a selection moment in the period of the α-helix, while a dominant signal at $\theta = 160° - 180°$ reflects a selection moment in the period of the β strand/sheet. We record the periodicity ($2\pi/\theta$) for which the selection moment achieves a maximum. If we know a structure for the protein family, we could compute the selection moments at periodicities matched with the protein structure. If no representative structure is available, the selection moment is computed for all blocks of contiguous residues of a preassigned length (say 24 residues, which is the average length of a transmembrane α-helix), and for each block, we derive a profile of the selection moment as a function of $\theta$. The maxima of this profile are then analyzed in tandem with predictions of the protein secondary structure.

To ascertain the correlation between selection moment and secondary structure, we derive selection moment profiles for different types of secondary structure elements, and note where the majority of peaks fall. We would expect the maximum selection moment to coincide with the period of the repetitive element. As an extension of the utility of the selection moment concept, we define the selection moment plot, which shows the mean selection moment as a function of mean conservation. Mean conservation is the average of the sum of the numerical conservation scores of the segment (identical to the value of the selection moment profile for an infinite period), and the mean selection moment is the corresponding ratio of the selection moment to the length of the segment. A high mean conservation and a low mean selection moment would be diagnostic of an interfacial structure that has a functional constraint to preserve both its lipid-facing and solvent-facing sides. Similar insights may be drawn with respect to the location of other types of structures in this plot.

Instead of the directional differential conservation described above, we may also expect to find a sequential differential conservation. The quantity of interest is then the difference in average conservation between the two halves, and larger differences point to a significance for the sequential differential conservation. It is predicted that the selection moment will be an important driving force in protein evolution, whereby it achieves a maximum trade-off between selection pressure and random variation. It is plausible that the evolution of most sequences tends to maximize their selection moment. Functional constraint is the deciding factor in the evolution of periodicity of mutability. However, unlike physical properties (like hydrophobicity) that can be studied for a single sequence, oscillations in mutability are limited by contemporaneity; it is clear they cannot be observed for a single sequence. Therefore our conclusions regard protein families.

The information derived from selection moments would be indispensable in critically evaluating the oscillations in physical properties themselves. For instance, if a high protein hydrophobic moment coincided with a high family selection moment, then the functional importance of the particular repetitive structure is reinforced. The permeation domain of an ion channel is a classic example of this: transmembrane α-helices flanking the pore of a $K^+$-channel evolved large conservation moments. In particular, the part of the inner helix facing the central pore is very conserved, but the part involved in packing with the outer helix is mainly variable (Palaniappan 2005). This emerges as a design feature in the architecture of ion channels. It is clear that selection pressure operates at the level of secondary structure to conserve the helical property of orientation of physical property. Selection moments might also play a crucial role in the specificity of oligomerisation of multimeric proteins.

The methodology described may be generalized with little modification to understand the directional evolutionary force acting on the surface of any given repetitive structure. Selection moment is a true essence concept, and it will be valuable in investigations of evolutionary mechanisms targeting important functions.